# Electrical Transport in High Quality Graphene *pnp* Junctions

Jairo Velasco Jr., Gang Liu, Wenzhong Bao, and Chun Ning Lau*
*To whom correspondence should be address. E-mail: lau@physics.ucr.edu

Department of Physics and Astronomy, University of California, Riverside, Riverside, CA 92521

**Abstract**

We fabricate and investigate high quality graphene devices with contactless, suspended top gates, and demonstrate formation of graphene *pnp* junctions with tunable polarity and doping levels. The device resistance displays distinct oscillations in the *npn* regime, arising from the Fabry-Perot interference of holes between the two *pn* interfaces. At high magnetic fields, we observe well-defined quantum Hall plateaus, which can be satisfactorily fit to theoretical calculations based on the aspect ratio of the device.

## I. Introduction

Graphene[1-3] is a two dimensional allotrope of carbon with a unique linear dispersion relation for low-lying excitations. Its gapless electronic band structure allows continuous tuning of charge carrier type and density by an electrostatic gate. Thus, one of the unique device configurations enabled by graphene is a dual-gated device, in which two or more gates are used to individually control charge density in different regions, realizing, for instance, *pnp* or *npn* junctions with *in situ* tunable junction polarity and doping levels[4-6]. Such *pnp* junctions have been demonstrated experimentally[7-14], offering unique platforms for investigation of novel phenomena such as Klein tunneling[15,16], particle collimation[4,17], anisotropic transmission[18] and Veselago lensing effects[19].

Here we report fabrication of high quality *pnp* junctions using suspended, contactless top gates. Distinct resistance oscillations in the bipolar region are observed, arising from Fabry-Perot interference of holes between the two *pn* interfaces. At high magnetic fields, device conductance display quantum Hall plateaus at fractional values of $e^2/h$, as a result of edge state equilibration. Using a recently available theory for rectangular device geometry[20], we can satisfactorily account for the two-terminal device conductance at uniform charge densities.

## 2. Device fabrication

Because graphene consists of only one atomic layer of carbon, deposition of top gate materials may introduce defects, dopants or additional scattering sites, thus reducing the device mobility. Previously, electron-beam resists[8,10] or alternating layers of $NO_2$, trimethylaluminum and $Al_2O_3$[9] have been used as the dielectrics for local gates. We have recently developed a multi-level lithography technique to fabricate "air-bridge"-styled top gates, in which a metallic bridge is suspended across part of the graphene sheet, with vacuum acting as the dielectric. A similar technique was also reported by ref. 12. Since graphene is only exposed to conventional lithographical resists and developers during fabrication, this technique minimizes the damage to the atomic layer. It is also compatible with post-fabrication annealing[21,22], which improves device mobility by removing resists residual and adsorbates on graphene surface, including those from the region directly under the top gate.

The substrate consists of a 300-nm $SiO_2$ layer grown over degenerately doped silicon, which can serve as a global back gate. Graphene sheets are deposited onto the substrate by standard micro mechanical exfoliation method[1], and we identify the number of layers in a given sheet by color contrast in an optical microscope and Raman spectroscopy[23]. To fabricate the air bridge, we employ a resist bilayer with different exposure and developing conditions, so as to create a temporary support for the suspended structure; this resist support is subsequently removed after double angle metal depositions at -45° and 45°. The details of fabrication are discussed in ref. 11. Finally, using standard electron beam lithography, we deposit electrodes, consisting of 10 nm of Ti and 80 nm of Al, on graphene. The height of the air gap depends on the thickness of resists, and is as small as 50 nm for our devices (Fig. 1a-b). To ensure suspension of the air bridge over distances as long as 12 μm, a critical point dryer is typically used during fabrication of the top gate and electrodes.

The suspended gates can also be fabricated at non-zero angles with respect to the source-drain electrodes (Fig. 1c-d), by careful control of the direction of metal evaporations. Such graphene devices with angled top gates were first proposed in ref. 15, as an experimental platform to observe Klein tunneling, in which the transmission coefficient of charges across high potential barriers strongly depends on the incident angle.

The most common failure mechanism of a suspended bridge is its collapse under sufficiently high voltages. Our previously fabricated device typically failed at critical voltages of ~40V-60V, due to the poor contacts between the vertical sidewalls and the horizontal bridge and electrodes. To overcome this deficiency, we perform an additional evaporation at 0º. Fig. 2a and 2b show two sets of suspended air bridges fabricated with two and three angle evaporations, respectively. Compared to Fig. 2a, the "joints" between different segments (outlined by dotted circles) in Fig. 2b are visibly strengthened.

To verify the mechanical robustness of these structures, we perform *in situ* SEM imaging while applying voltages to the top gate. As shown by the images in Fig. 2c-e, the air bridge remains suspended and undeformed under voltages of 70V and 100V, and ultimately fails at 110V. This surprisingly high critical voltage demonstrates significant improvement over our previous top gate structures.

## 3. Conductance of a *pnp* junction at *B*=0

Transport measurements on the graphene devices are performed at 260mK in a He$^3$ fridge using standard lock-in techniques. By varying voltages applied to the back gate (that controls the charge density and type in the entire device) and to the top gate (that controls charges directly under it), a graphene *pnp* junction can be created. A typical data set is shown in Figure 3a, plotting the 4-terminal device resistance $R$ (color) as functions of $V_{bg}$ (vertical axis) and top gate voltage $V_{tg}$ (horizontal axis). The source-drain separation $L$ of the device is 3.5 μm, and width $W$ is 1 μm. The top gate is suspended $d$~100 nm above the center segment of graphene, with a length $L_{tg}$≈0.5 μm. In the non-top-gated or "bare" regions of the device, the charge density $n_1$ are modulated by $V_{bg}$ only,

$$n_1 = C_{bg}(V_{bg} - V_{D,bg})/e \qquad (1a)$$

where $e$ is the electron's charge, $C_{bg}$ is the capacitance per unit area between graphene and the back gate, and $V_{D,bg}$ is the Dirac point of the bare region of the device, which may be non-zero due to doping by contaminants. From Fig. 3a, $V_{D,bg}$ ~ 3.5V, at which device resistance reaches a maximum; since these regions account for 85% of the device area, their response dominate the device resistance, yielding the horizontal green-red band.

For the top-gated or "covered" region, the charge density $n_2$ are modulated by both $V_{bg}$ and $V_{tg}$,

$$n_2 = n_1 + C_{tg}(V_{tg} - V_{D,tg})/e \qquad (1b)$$

At $n_1$ =0, the resistance maximum occurs at the Dirac point of the top-gated region, $V_{D,tg}$ ~ 18V. The diagonal cyan line, indicating a local resistance maximum, corresponds to the charge neutrality point of the region directly under the top gate, *i.e.* $n_2$=0. Its slope thus yields the capacitance or gate coupling ratio, $\eta = C_{tg}/C_{bg}$, measured from the figure to be ~0.78. This is in good agreement with the value estimated from geometrical considerations, $\frac{C_{tg}}{C_{bg}} = \frac{\varepsilon_{tg}}{\varepsilon_{bg}} \frac{d_{bg}}{d_{tg}} \approx (1/3.9)(300/100) = 0.77$, where $\varepsilon$ is the dielectric constant of the gate, and $d$ is the device-gate separation. Taken together, the horizontal band of $n_1$=0, together with the cyan diagonal line of $n_2$=0, partition the data in Fig. 3a into four regions, with different doping combinations, thus demonstrating a graphene *pnp* junction with *in situ* tunable polarity and doping levels.

We now focus exclusively on the upper left region of Fig. 3a, *i.e.* when the junction is in the *npn* regime. Compared with the neighboring unipolar (*pp'p* or *nn'n*) regions, the junction resistance is significantly higher, as expected at the boundary of a *pn* junction. More interestingly, we observe resistance oscillations as a function of both $V_{bg}$ and $V_{tg}$, as indicated by the arrows in Fig. 3a. Notably, these oscillations are not found in the unipolar regions. Such oscillations were first reported by ref. 13, and arise from Fabry-Perot interference of the charges between the two *p-n* interfaces. Thus, the holes in the top-gated region are multiply reflected between the two interfaces, interfering to give rise to standing waves, similar to those observed in carbon nanotubes[24] or standard graphene devices[25]. Modulations in $n_2$ change the Fermi wavelength of the charge carriers, hence altering the interference patterns and giving rise to the resistance oscillations.

To analyze these oscillations in detail, we replot the data in Fig. 3a in terms of $n_1$ and $n_2$. Assuming a parallel plate geometry between the gate and the device, $C_{bg}/e \approx 7.19 \times 10^{10}$ cm$^{-2}$. However, from quantum Hall measurements, we estimate the effective capacitance to be $\sim C_{bg}/e \approx 6.51 \times 10^{10}$ cm$^{-2}$ (see discussion in the next section). This small discrepancy may be attributed to a slightly thicker SiO$_2$ layer, slightly smaller $\varepsilon_{bg}$, or additional screening by the electrodes. Using this value, we have

$$n_1 = 6.5 \times 10^{10}(V_{bg} - V_{D,bg}) \text{ cm}^{-2} \qquad (2a)$$

and

$$n_2 = 6.5 \times 10^{10}[(V_{bg} - V_{D,bg}) + \eta(V_{tg} - V_{D,tg})] \text{ cm}^{-2} \ . \qquad (2b)$$

The new plot is shown in Fig. 3b. The color scale is adjusted to accentuate the resistance oscillations, which appear as fringes fanning out from the Dirac point at $n_1 = n_2 = 0$. Fig. 3c shows the device resistance *vs.* $n_2$ at $n_1 = 1.3 \times 10^{12}$ cm$^{-2}$, displaying clear oscillations.

Within the Fabry-Perot model, the resistance peaks correspond to minima in the overall transmission coefficient; the peak separation can be approximated by the condition $k_F(2L) = 2\pi$, *i.e.* a charge accumulates a phase shift of $2\pi$ after completing a roundtrip $2L_c$ in the cavity. Here $k_F$ is the Fermi wave vector of the charges, and $L_c$ is the length of the Fabry Perot cavity. Under the top gate, $k_{F2} = \sqrt{\pi n_2}$, so the spacing between successive peaks is estimated to be

$$\Delta n_2 \approx \frac{4\sqrt{\pi n_2}}{L_c} \qquad (3)$$

In Fig. 3d, we plot the measured peak spacing for the curve shown in Fig. 3c against $\sqrt{n_2}$. The data points fall approximately on a straight line. The best linear fit yields a line with a slope 0.95 x10$^5$/cm, corresponding to $L_c = 740$ nm from Eq. (3). This agrees with the value estimated from electrostatics, $L_c = L_{tg} + 2d$, as the electric field induced by the top gate on the device is expected to extend by a distance $\sim d$ away from either edge.

Finally, we note that the device in ref. 13 had extremely narrow gates $L_{tg} < \sim 20$ nm. In comparison, our top gate spans a much larger distance, $L_{tg} \sim 500$ nm. Thus, the observation of clear Fabry-Perot interference patterns underscores the high quality of our *pnp* graphene devices.

**4. Conductance of a *pnp* junction at *B*=8T**

In high magnetic fields, the cyclotron orbits of charges coalesce to form Landau levels (LLs). For graphene with uniform charge densities, the energies of the LLs are given by

$E_N = \text{sgn}(N)\sqrt{2e\hbar v_F^2 |N|B}$, and the Hall conductivity is $\sigma_{xy} = 4(N+\tfrac{1}{2})\frac{e^2}{h}$, where $N$ is an integer denoting the LL index, $e$ is the electron charge, $v_F$ is the Fermi velocity of charges in graphene, $h$ is the Planck's constant. The factor of 4 arises from the spin and sublattice degeneracies. A unique signature of graphene's linear dispersion relation is the presence of the zeroth LL that is shared equally by electrons and holes, leading to quantum Hall plateaus that are quantized at half-integer values of $4e^2/h$ [1, 3].

For a graphene device with dual gates, the situations are complicated by the presence of regions with different filling factors, or the co-existence of n- and p-doped regions that result in counter-propagating edge states. The two-terminal conductance of the junction depends on the relative values of $n_1$ and $n_2$, and can have fractional values of $e^2/h$. A simple model is provided in ref. 26, assuming perfect edge state equilibration at the interfaces between different regions: for a unipolar junction ($n_1 n_2 > 0$) with $|n_1| \geq |n_2|$, the non-top-gated regions act as reflectionless contacts to the center region, yielding a device conductance

$$G = e^2/h |v_2| \qquad (4)$$

where $v_1$ and $v_2$ are the filling factor in the bare and top-gated regions, respectively, given by $v = \frac{nh}{Be}$, where $B$ is the applied magnetic field. If instead $|n_2| > |n_1|$, the conductance is

$$G = \frac{e^2}{h}\left(\frac{1}{|v_1|} - \frac{1}{|v_2|} + \frac{1}{|v_1|}\right)^{-1} \qquad (5)$$

For a bipolar junction ($n_1 n_2 < 0$), the device behaves simply as three resistors in series,

$$G = \frac{e^2}{h}\left(\frac{1}{|v_1|} + \frac{1}{|v_2|} + \frac{1}{|v_1|}\right)^{-1} \qquad (6)$$

From Eqs. (4-6), the conductance of a graphene *pnp* junctions display plateaus at fractional multiples of $e^2/h$. We emphasize that these fractional-valued plateaus are not related to the fractional quantum Hall effect; rather, they arise from the inhomogeneous filling factors within the device.

To observe these plateaus, we measure the two-terminal conductance $G$ of the device as functions of $V_{bg}$ and $V_{tg}$ at $B=8$T. The data are shown in Fig. 4a, plotting $G$ (color) vs. $n_1$ and $n_2$, which are calculated from gate voltages using Eqs. (1) and (2). The conversion factor $C_{bg}/e$ is obtained by noting that, measured from the global Dirac point ($n_1=n_2=0$), center of the first finite density plateau should occur at $v=2$, or $n=3.9\times10^{11}$ cm$^{-2}$ at 8T; the corresponding gate voltage difference is 6 V, yielding an effective $C_{bg}/e \sim 6.51\times10^{10}$ cm$^{-2}$V$^{-1}$. This value is ~9% lower than that estimated from a simple parallel plate capacitor model, and is used for all plots in Fig. 3 and 4.

The data in Fig. 4a exhibit regular rectangular patterns, which arise from the filling of additional LLs as $n_1$ and $n_2$ are modulated. A line trace of $G(n_2)$ at $v_1=2$ is shown in Fig. 4b, with equivalent values of $v_2$ labeled on the top axis. As $v_2$ varies from -2, 2 to 6, conductance plateaus with values of ~0.67, 2 and 1.2 $e^2/h$ are observed, in excellent agreement with those obtained using Eqs. (6-8). The solid line in Fig. 4c plots $G(v)$ for uniform charge densities over the entire graphene sheet, *i.e.*, along the diagonal dotted line $n_2=n_1$ in Fig. 4a. The $v=2$ plateau is well-developed, indicating relatively small amount of disorder.

We now focus on the small conductance dips in Fig. 4c at $v\sim3$ and $v\sim7$, which are not expected to be present for a square device with $L=W$. Indeed, the two-terminal conductance of a conducting *square* includes both longitudinal and Hall conductivity signals, $G = \sqrt{\sigma_{xx}^2 + \sigma_{xy}^2}$, so

$G(\nu)$ appears as stepwise plateaus that increases monotonously for $\nu>0$. However, for other device geometries the behaviors are more complicated. Depending on the aspect ratio of the device, the device conductance displays local conductance peaks or dips between the plateaus; if the device has significant Landau level broadening, the conductance will no longer be quantized at integer values of 2, 6, 10… $e^2/h$. This was studied in detail in ref. 20, using an effective medium approach that yields a semicircle relation between $\sigma_{xx}$ and $\sigma_{xy}$. To quantitatively examine the agreement between the data and the theory, we model the longitudinal conductivity as a Gaussian centered at a LL, $\delta_N \sigma_{xx}(\nu) = 2e^{-[\nu-1/2(\nu_N+\nu_{N+1})]^2/\Gamma}$ in units of $e^2/h$. Here $\nu_N = 4(N+\frac{1}{2})\frac{|B|e}{h}$ are the incompressible densities corresponding to the $N$th LL, and $\Gamma$ describes the width of the Gaussian distribution. Following the procedures outlined in ref. 20, and using a fitting parameter $\Gamma$=0.67, we calculate $G(\nu)$ for our rectangular device with aspect ratio $L/W$=3.5. The resultant curve is shown as the dotted line in Fig. 4c. The agreement with data is satisfactory at smaller values of $\nu$, but deviates for $\nu>6$. This is quite reasonable, since the energetic difference between Landau levels decrease for higher levels. Moreover, the value of $\Gamma$=0.67 obtained from the fitting is relatively small, again underscoring the high junction quality.

Finally, we note that the model leading to Eq.s (4)-(6) is based on the single particle picture and assumes perfect edge state equilibration at the interfaces between different regions. The excellent agreement between our experimental results and model validates this assumption. On the other hand, if the electrical transport is fully coherent, one expects to observe universal conductance fluctuations (UCF) instead of well-defined plateaus. The exact origin of such mode-mixing mechanism and suppression of UCF is not clear, but is likely related to the presence of disorder, and/or dephasing due to coupling of the edge states to localized states in the bulk. Thus, an interesting future direction to explore is the mode-mixing mechanism (or the lack thereof) by studying ultra-clean *pnp* junctions.

## 5. Conclusion

Using suspended top gates, we are able to fabricate high quality *npn* junctions, which display Fabry-Perot resistance oscillations within a small cavity formed by the *pn* interfaces. In high magnetic fields, well-developed quantum Hall plateaus are observed, and the behavior can be quantitatively described by theoretical predictions for rectangular device geometry. In the long term, this technique may be important for study of transport of Cooper pairs[27-29] or spins[30,31] through *pn* junctions, or the experimental realization of on-chip electronic lenses[19], which require extremely clean graphene devices.

**Acknowledgement**

We thank Marc Bockrath, Misha Fogler and Shan-Wen Tsai for discussions. This work is supported in part by NSF/CAREER DMR/0748910, ONR N00014-09-1-0724 and UC Lab Fees Research Program 09-LR-06-117702-BASD.

**References**
1. K. S. Novoselov, A. K. Geim, S. V. Morozov, D. Jiang, M. I. Katsnelson, I. V. Grigorieva, S. V. Dubonos and A. A. Firsov, *Nature*, 2005, **438**, 197-200.
2. K. S. Novoselov, A. K. Geim, S. V. Morozov, D. Jiang, Y. Zhang, S. V. Dubonos, I. V. Grigorieva and A. A. Firsov, *Science*, 2004, **306**, 666-669.
3. Y. B. Zhang, Y. W. Tan, H. L. Stormer and P. Kim, *Nature*, 2005, **438**, 201-204.
4. V. V. Cheianov and V. I. Fal'ko, *Phys. Rev. B*, 2006, **74**, 041403.
5. M. M. Fogler, L. I. Glazman, D. S. Novikov and B. I. Shklovskii, *Phys. Rev. B*, 2008, **77**, 075420.
6. L. M. Zhang and M. M. Fogler, *Phys. Rev. Lett.*, 2008, **100**, 116804.
7. B. Ozyilmaz, P. Jarillo-Herrero, D. Efetov, D. A. Abanin, L. S. Levitov and P. Kim, *Phys. Rev. Lett.*, 2007, **99**, 166804.
8. B. Huard, J. A. Sulpizio, N. Stander, K. Todd, B. Yang and D. Goldhaber-Gordon, *Phys. Rev. Lett.*, 2007, **98**, 236803.
9. J. R. Williams, L. DiCarlo and C. M. Marcus, *Science*, 2007, **317**, 638.
10. J. B. Oostinga, H. B. Heersche, X. L. Liu, A. F. Morpurgo and L. M. K. Vandersypen, *Nature Materials*, 2008, **7**, 151-157.
11. G. Liu, J. Velasco, W. Z. Bao and C. N. Lau, *Appl. Phys. Lett.*, 2008, **92**, 203103.
12. R. V. Gorbachev, A. S. Mayorov, A. K. Savchenko, D. W. Horsell and F. Guinea, *Nano Lett.*, 2008, **8**, 1995-1999.
13. A. F. Young and P. Kim, *Nature Phys.*, 2009, **5**, 222-226.
14. T. Lohmann, K. v. Klitzing and J. H. Smet, *arXiv:0903.5430v1*, 2009.
15. M. I. Katsnelson, K. S. Novoselov and A. K. Geim, *Nature Physics*, 2006, **2**, 620-625.
16. A. V. Shytov, M. S. Rudner and L. S. Levitov, *Phys. Rev. Lett.*, 2008, **101**, 156804.
17. C. H. Park and S. G. Louie, *Nano Lett.*, 2008, **8**, 2200-2203.
18. C. H. Park, L. Yang, Y. W. Son, M. L. Cohen and S. G. Louie, *Nature Phys.*, 2008, **4**, 213-217.
19. V. V. Cheianov, V. Fal'ko and B. L. Altshuler, *Science*, 2007, **315**, 1252-1255.
20. D. A. Abanin and L. S. Levitov, *Phys. Rev. B*, 2008, **78**, 035416.
21. J. Moser, A. Barreiro and A. Bachtold, *Appl. Phys. Lett.*, 2007, **91**, 163513.
22. J. H. Chen, C. Jang, S. D. Xiao, M. Ishigami and M. S. Fuhrer, *Nature Nanotechnol.*, 2008, **3**, 206-209.
23. A. C. Ferrari, J. C. Meyer, V. Scardaci, C. Casiraghi, M. Lazzeri, F. Mauri, S. Piscanec, D. Jiang, K. S. Novoselov, S. Roth and A. K. Geim, *Phys. Rev. Lett.*, 2006, **97**, 187401.
24. W. J. Liang, M. Bockrath, D. Bozovic, J. H. Hafner, M. Tinkham and H. Park, *Nature*, 2001, **411**, 665-669.
25. F. Miao, S. Wijeratne, Y. Zhang, U. Coskun, W. Bao and C. N. Lau, *Science*, 2007, **317**, 1530-1533.
26. D. A. Abanin and L. S. Levitov, *Science*, 2007, **317**, 641-643.
27. H. B. Heersche, P. Jarillo-Herrero, J. B. Oostinga, L. M. K. Vandersypen and A. F. Morpurgo, *Nature*, 2007, **446**, 56-59.
28. X. Du, I. Skachko and E. Y. Andrei, *Phys. Rev. B*, 2008, **77**, 184507.
29. F. Miao, W. Bao, H. Zhang and C. N. Lau, *Solid State Commun.*, 2009, **149**, 1046.
30. N. Tombros, C. Jozsa, M. Popinciuc, H. T. Jonkman and B. J. van Wees, *Nature*, 2007, **448**, 571-574.

31. W. Han, W. H. Wang, K. Pi, K. M. McCreary, W. Bao, Y. Li, F. Miao, C. N. Lau and R. K. Kawakami, *Phys. Rev. Lett.*, 2009, **102**, 137205.

Figure 1. SEM images of suspended top gates. (a). A completed graphene device with a top gate suspended ~ 100 nm above the substrate. (b). Image of the top gate suspended at 50 nm above the substrate. (c-d). A graphene device with a suspended top gate that is not parallel to the electrodes. The images are taken at tilt angles of 0º and 70º, respectively.

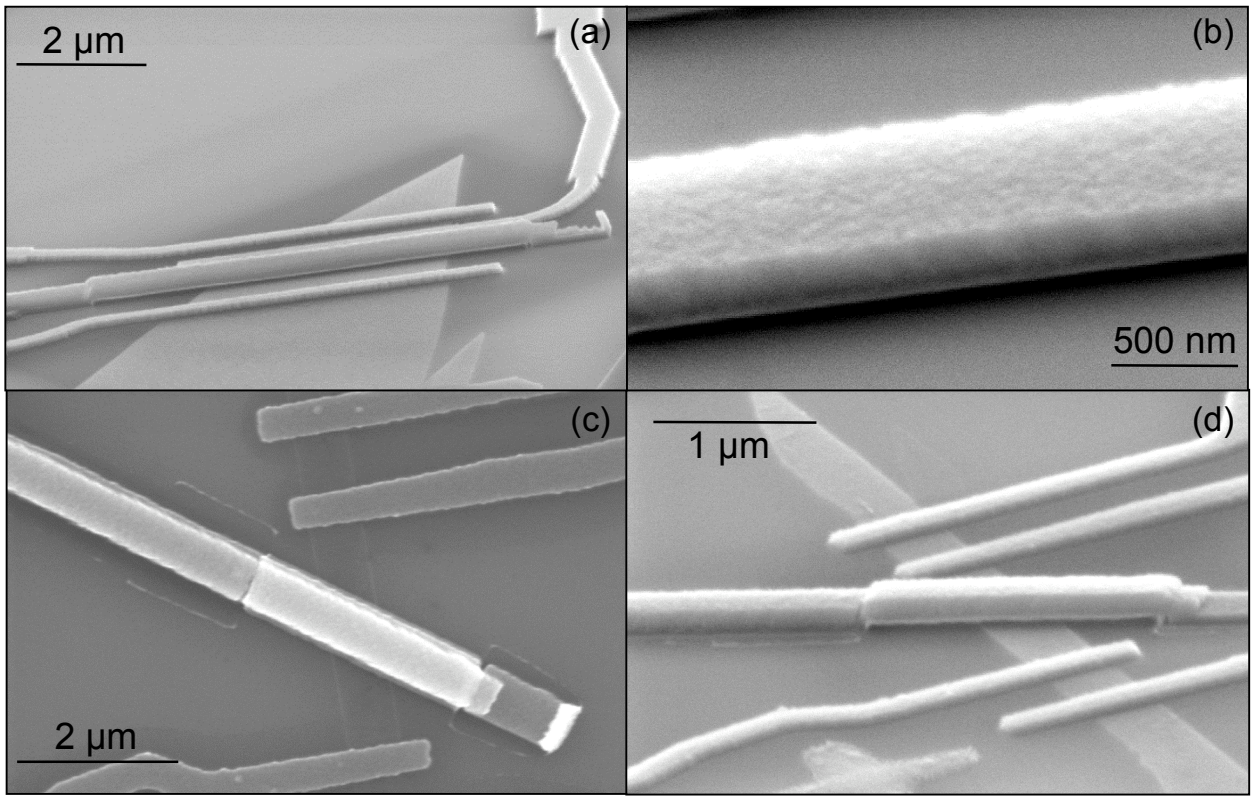

Figure 2. (a). SEM image of suspended test structures fabricated with double angle evaporations (at -45º and 45º). The red circles indicate the weak points of the structure. (b). Same as (a), but fabricated with triple angle evaporations (at -45º, 45º and 0º). The weak points are visibly reinforced. (c-e). SEM images of a suspended top gate under applied voltages of 70V, 100V and 110V, respectively.

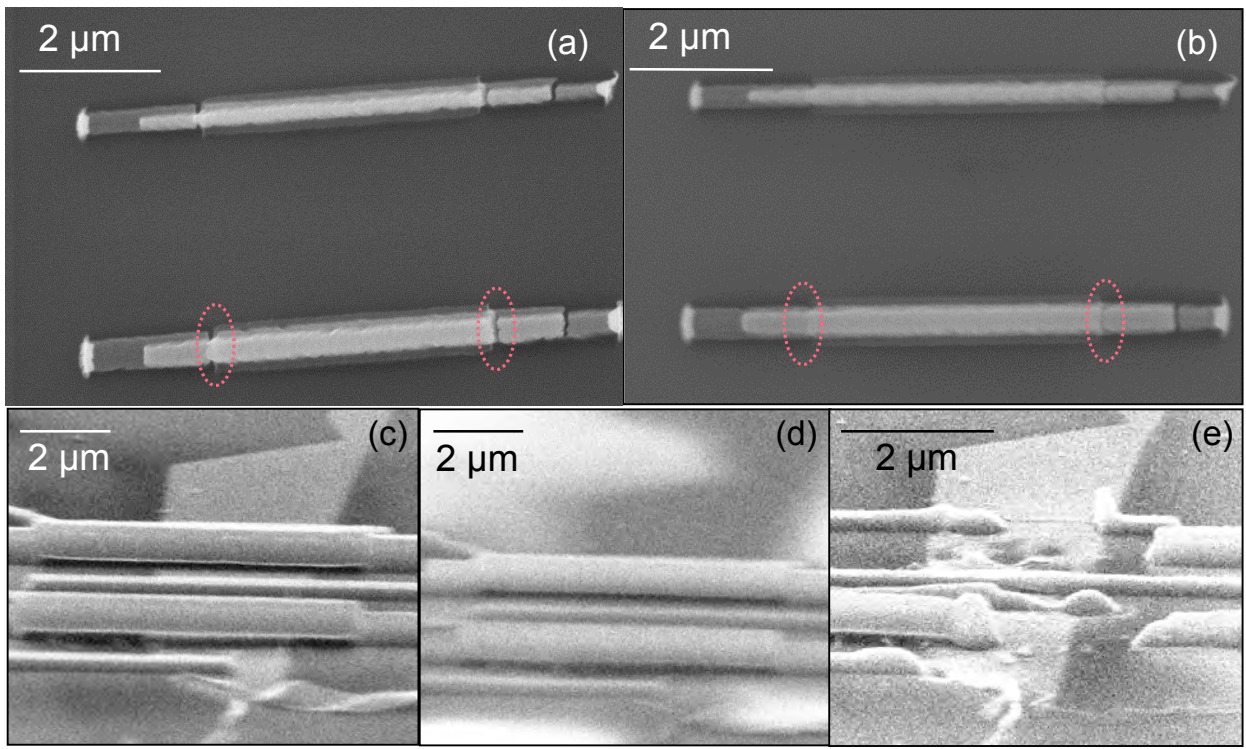

Figure 3. Data in zero magnetic field. (a). Four-terminal device resistance as functions of $V_{bg}$ and $V_{tg}$. The arrows indicate oscillations in the *npn* region. (b). Same data as (a), but plotted against $n_2$ and $n_1$. (c). Line trace along the dotted line in (b), showing resistance oscillation as a function of $n_2$. (d). The peak spacing $\Delta n_2$ vs. $\sqrt{n_2}$. The line represents a linear fit to the data.

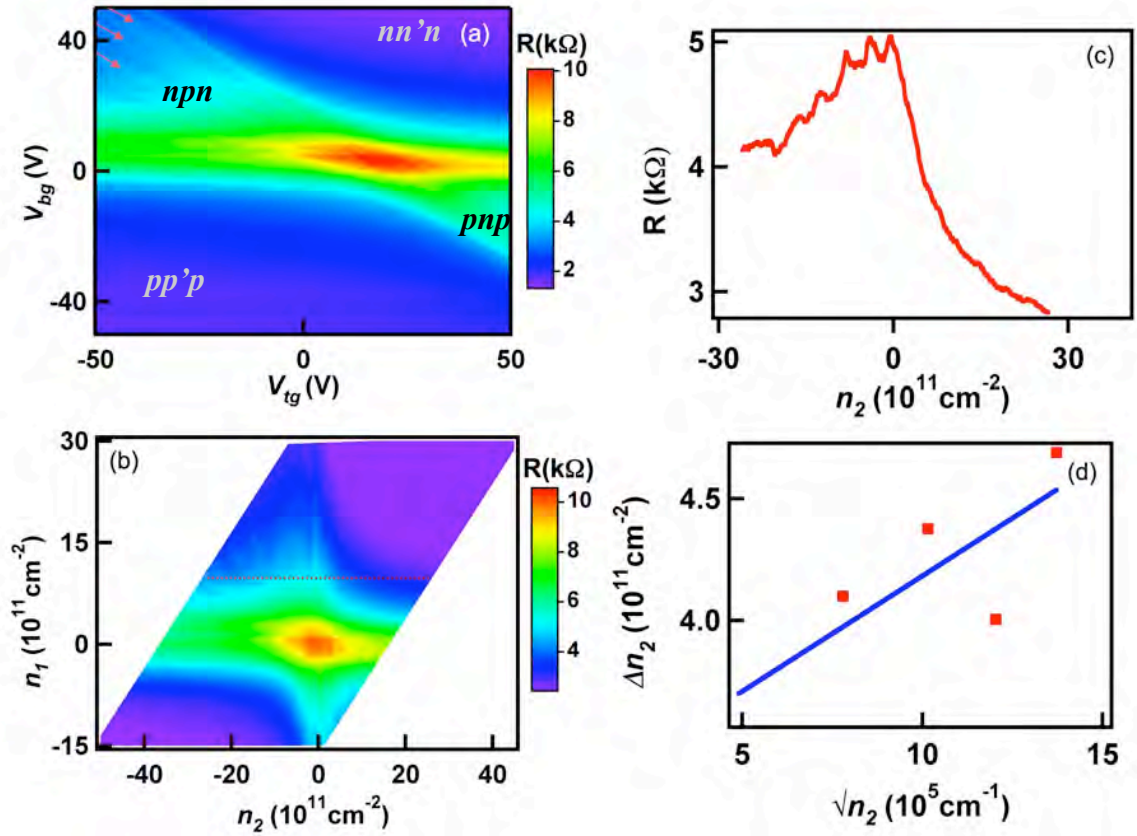

Figure 4. Data at $B=8T$. (a). Two-terminal device conductance $G$ vs. $n_1$ and $n_2$. (b). Line trace along the green dotted line $v_1=2$ in (a). (c). The solid red curve is a line trace taken along the magenta dotted line $n_2= n_1$ in (a). The dotted line is a theoretical curve calculated using the expressions in ref. 20, $L/W=3.5$ and $\Gamma=0.67$, where G is a fitting parameter related to the Laudau level broadening.

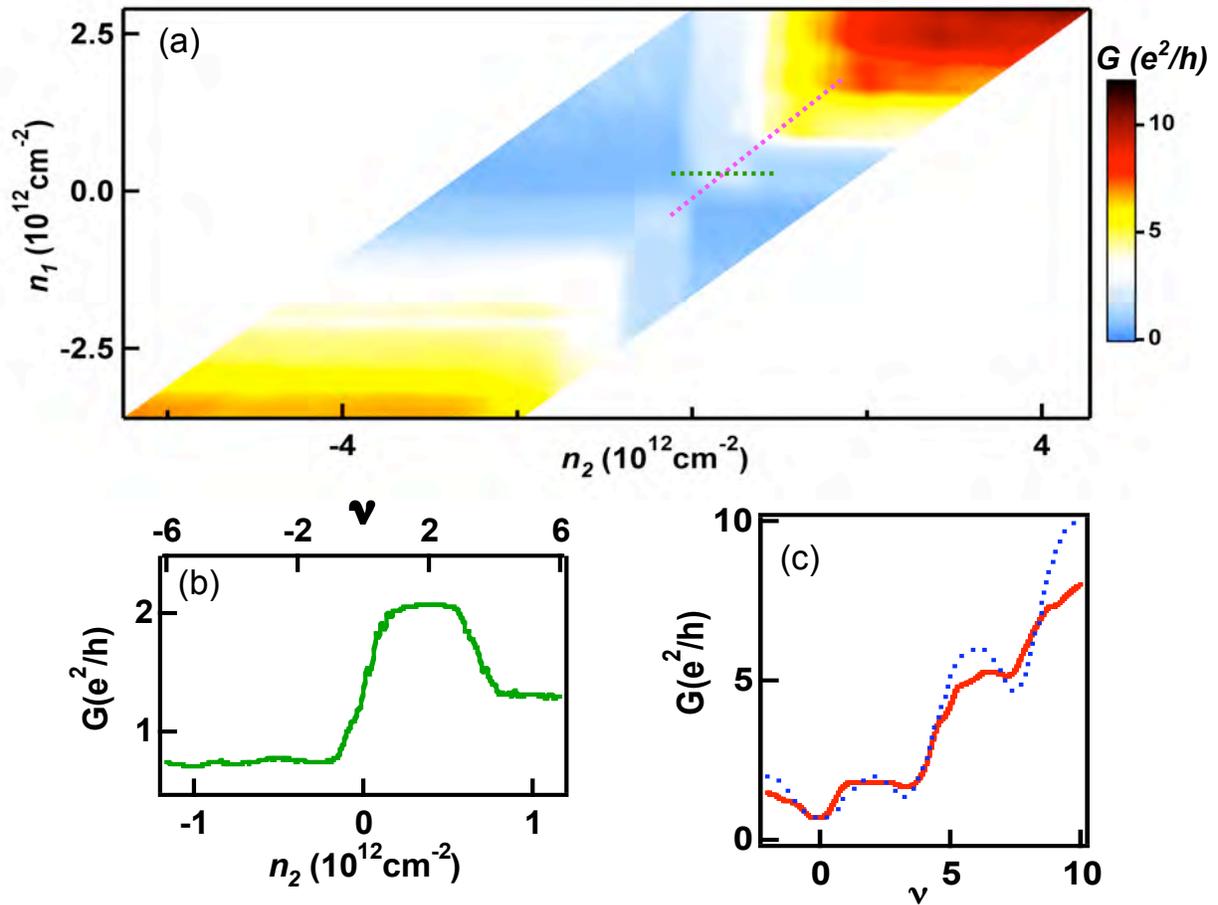